\documentclass[fleqn,twoside]{article}
\usepackage{espcrc2,cite,graphicx,amsmath,amsfonts,amssymb,mathrsfs}
\usepackage[figuresright]{rotating}

\newcommand{\ie}{{\it i.e.}}

\newcommand{\eg}{{\it e.g.}}

\newcommand{\cf}{{\it cf.}}

\newcommand{\eq}{Eq.}

\newcommand{\fig}{Fig.}

\newcommand{\Ref}{Ref.}
\newcommand{\Refs}{Refs.}

\newcommand{\App}{App.}

\begin{document}

\title{
\vspace*{-3cm}
\begin{flushright}
{\small TUM-HEP-449/01}
\end{flushright}
\vspace*{0.5cm}
{\bf Effects of random matter density fluctuations on the neutrino oscillation
transition probabilities in the Earth}}

\author{{\large Bj{\"o}rn Jacobsson}\address[KTH]{{\it Division of
Mathematical Physics, Department of Physics, Royal Institute of
Technology - Stockholm Center for Physics, Astronomy, and
Biotechnology, 106 91 Stockholm, Sweden}}\thanks{E-mail: {\tt
bjorn@theophys.kth.se}},
{\large Tommy Ohlsson}\address[TUM]{{\it Institut f{\"u}r Theoretische
Physik, Physik-Department, Technische Universit{\"a}t M{\"u}nchen,
James-Franck-Stra{\ss}e, 85748 Garching bei M{\"u}nchen,
Germany}}\thanks{E-mail: {\tt tohlsson@ph.tum.de}},
{\large H{\aa}kan Snellman}\addressmark[KTH]\thanks{E-mail: {\tt
snell@theophys.kth.se}},
{\large Walter Winter}\addressmark[TUM]\thanks{E-mail: {\tt
wwinter@ph.tum.de}}}
     
\begin{abstract}
\noindent {\bf Abstract} 
\vspace{2.5mm}

In this paper, we investigate the effects of random fluctuations of
the Earth matter density for long baselines on the neutrino
oscillation transition probabilities. We especially identify relevant
parameters characterizing the matter density noise and calculate their
effects by averaging over statistical ensembles of a large number of matter
density profiles.
For energies and baselines appropriate to neutrino factories, 
absolute errors on the relevant appearance probabilities 
are at the level of $|\Delta P_{\alpha \beta}| \sim 10^{-4}$
(with perhaps $|\Delta P_{\mu e}|/P_{\mu e} \sim 1\%$ 
for neutrinos), whereby a modest improvement in understanding 
of the geophysical data should render such effects unimportant.

\vspace*{0.2cm}
\noindent {\it PACS:} 14.60.Lm, 13.15.+g, 91.35.-x, 23.40.Bw \\
\noindent {\it Key words:} Neutrino oscillations, Matter effects, Earth's
matter density profile, Long baseline neutrino experiments
\end{abstract}

\maketitle

\section{Introduction}

The effects of matter on neutrino oscillations
\cite{mikh85,mikh86,wolf78} in the Earth have been investigated in
various contexts and with several models
\cite{Ermilova:1986ph,Baltz:1987hn,Nicolaidis:1988fe,Krastev:1988yu,Kuo:1989qe,Krastev:1989ix,Minorikawa:1990ip,Petcov:1998su,Akhmedov:1998ui,Akhmedov:1998xq,Chizhov:1999az,Chizhov:1999he,Akhmedov:1999ty,Freund:1999vc,Ohlsson:1999xb,Ohlsson:1999um,Freund:1999gy,Mocioiu:2000st,Freund:2000ti,Dick:2000ce,Akhmedov:2000js,Ota:2000hf,Takahashi:2000it,Ohlsson:2001et,Ohlsson:2001au,Bernabeu:2001xn,Takahashi:2001dc}.
It is now well-known that the matter density can significantly change the
reconstructed neutrino energy spectrum produced by long baseline
neutrino experiments, such as by neutrino factories
\cite{Geer:1998iz,Barger:1999fs,Freund:1999gy}. For most calculations the
Preliminary Reference Earth Model (PREM) density profile \cite{dzie81} has
been used, which is obtained from geophysical seismic wave measurements (see,
\eg, \Refs~\cite{aki80,lay95,shea99} for information about the
structure of the Earth's interior). Furthermore, small errors in
the PREM matter density with up to 5\% amplitude have been found and
documented by many geophysics groups (for a summary, see, \eg,
\Ref~\cite{Pana}). Note, however, that the Earth's matter density
distribution is not directly observable from seismological
data \cite{bull75,kenn98}. In this paper, we will discuss how these
fluctuations in the Earth matter density affect the neutrino
oscillation transition probabilities.

It has been found that for short baselines matter effects are small
\cite{Akhmedov:2000cs}. Thus, any fluctuations could be treated
as second order effects and can therefore be neglected. For long baselines,
however, the fluctuations may be significant. It was noted in
\Ref~\cite{Shan:2001br} that this effect can be important as an
additional uncertainty in the determination of the CP phase $\delta_{CP}$,
especially for certain values of $\delta_{CP}$. To estimate this effect, the
authors used a logarithmic distribution with a certain length scale and
amplitude with a path integral method for the numerical evaluation. On
the other hand, it was shown in \Ref~\cite{Ohlsson:2001ck}, using a
perturbation theoretical approach, that fluctuations with small
amplitudes on length scales much shorter than the oscillation length
in matter average out and give no net effect at all. In this paper, we
are interested in the errors on the transition probabilities as
functions of the length scale and amplitude of the matter density
fluctuations. We will, in particular, focus on the errors on the
appearance probability of electron neutrinos and electron
antineutrinos at typical neutrino factory energies, since matter density
noise effects could be rather substantial in the determination of the 
CP phase and matter effects are largest in the appearance channel.

\section{The effect of a matter density perturbation}

Before we come to modeling the fluctuations in the Earth matter
density profile, let us study the effect of a single perturbation in the
matter density. In any quantum mechanical system, described by a
Schr{\"o}dinger equation, the impact on a free particle's motion of a
potential depends on its length scale as well as its 
amplitude. When the length scale of the potential is much shorter than
the characteristic wave length of the incident particle, the
particle's wave function will be unable to resolve the exact spatial
structure of this potential. It is then possible to replace the 
potential by a $\delta$-distribution with an amplitude equal to the
integral of the original potential.
Similarly, in neutrino oscillations, described by a Schr{\"o}dinger equation,
a perturbation on a length scale much shorter than the oscillation
length in matter could be replaced by a $\delta$-distribution.
\begin{figure*}[ht!] 
\begin{center} 
\includegraphics[width=7cm]{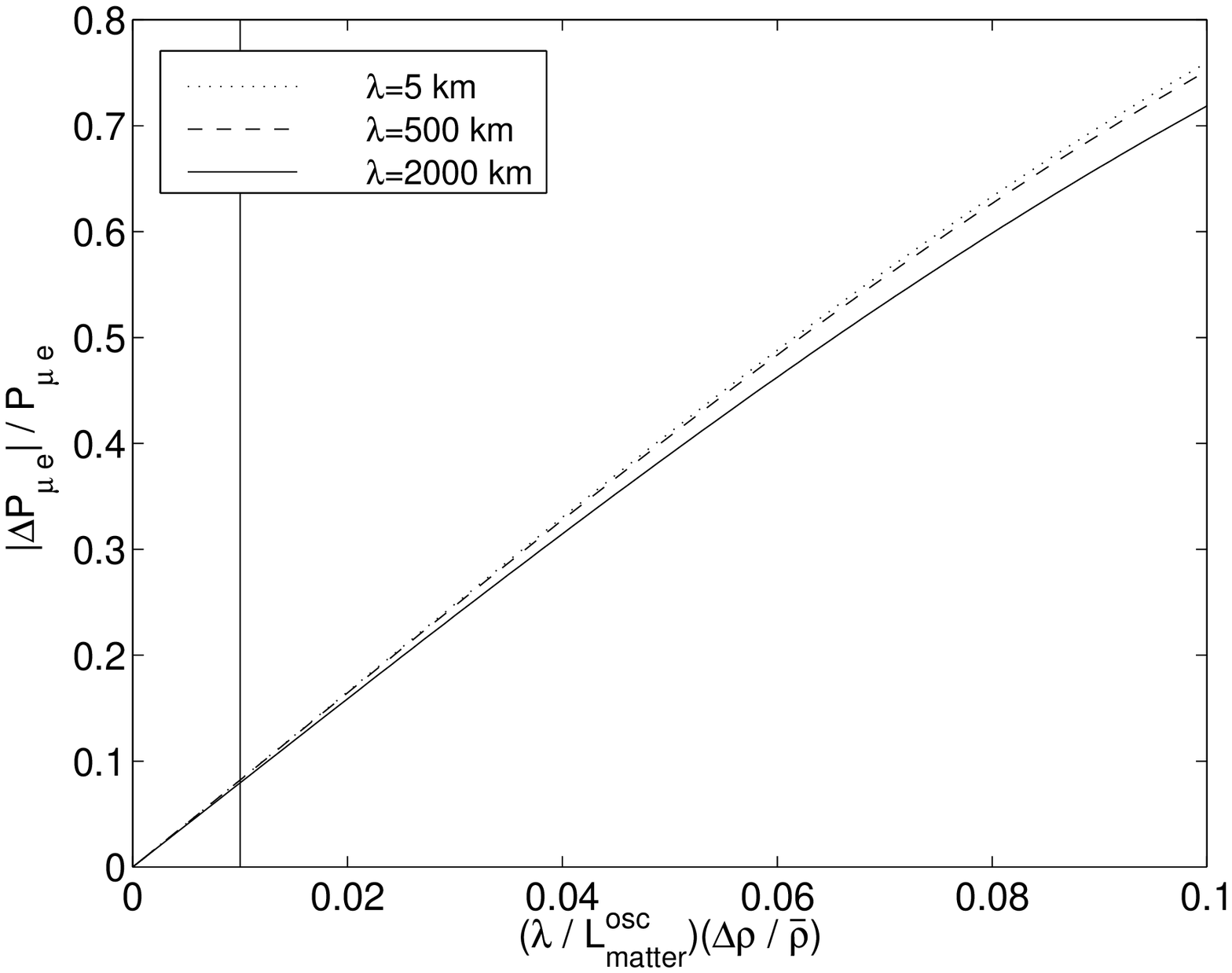}
\hspace*{1cm} 
\includegraphics[width=7cm]{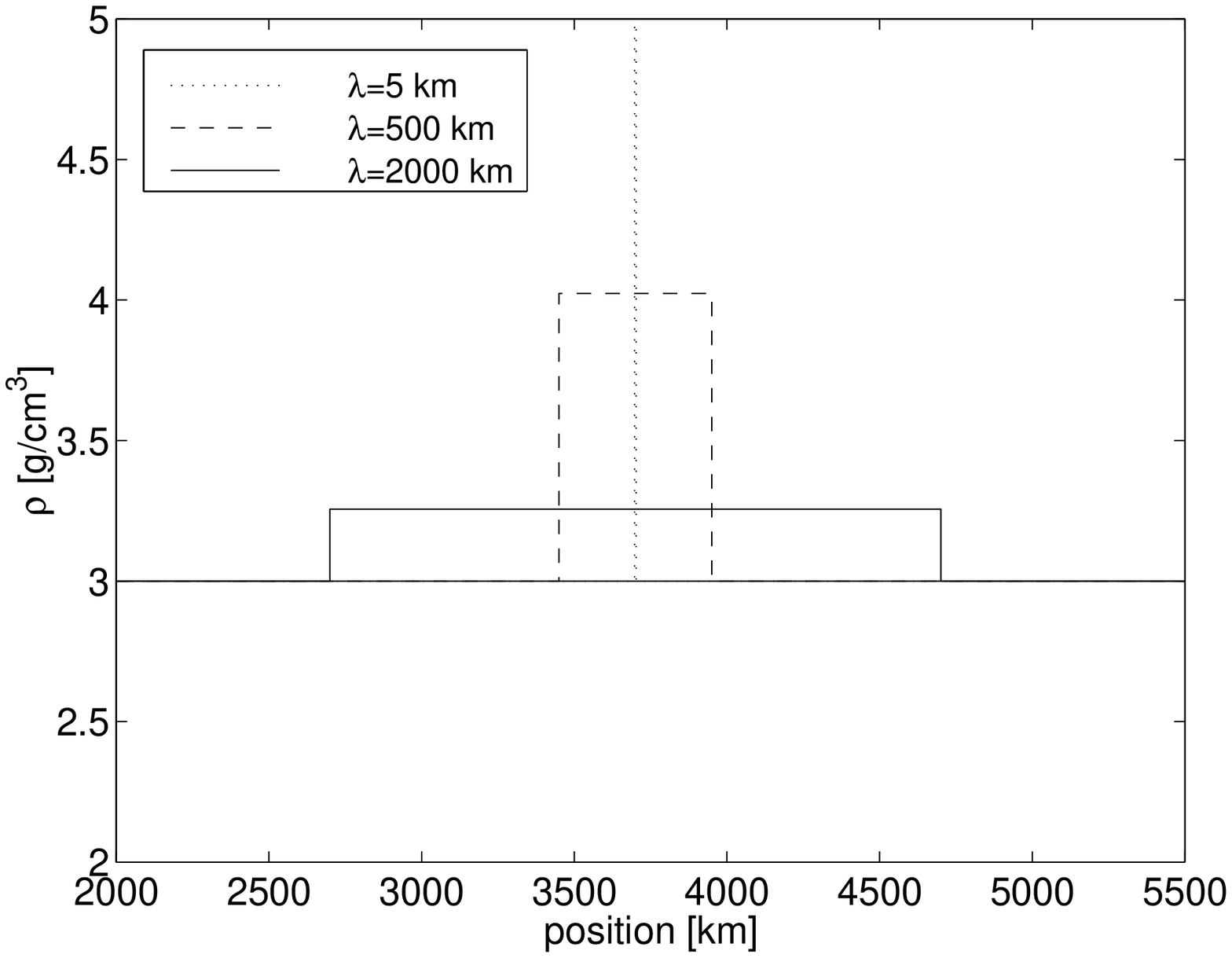} 
\end{center} 
\vspace*{-1cm}
\caption{\label{pert} The plot on the left-hand side shows the relative
error $| \Delta P_{\mu e} |/P_{\mu e}^{\mathrm{ref}}$ in the appearance
probability of the unperturbed profile $P_{\mu e}^{\mathrm{ref}}$ with a
rectangular matter density perturbation of length $\lambda$ and amplitude
$\Delta \rho \equiv \rho-\bar{\rho}$, where $\bar{\rho} = 3 \,
\mathrm{g/cm^3}$ is the average matter density. The perturbation is assumed
to be centered at a baseline of length $L=7400 \, \mathrm{km}$. The relative
error is plotted as a function of the product of the relative length scale
$\lambda/L^{\mathrm{osc}}_{\mathrm{matter}}$ and the relative density
contrast $\Delta \rho/\bar{\rho}$. Here
$L^{\mathrm{osc}}_{\mathrm{matter}} \simeq 17000 \, \mathrm{km}$ is
the oscillation length in matter, determined by the average matter
density $\bar{\rho}$ and the leading neutrino oscillation parameters
$\Delta m_{32}^2$ and $\theta_{13}$ for a typical neutrino factory 
maximum energy chosen to be $E = 30 \, \mathrm{GeV}$. The plot on the
right-hand side shows this density perturbation for a fixed value of the area
spanned by the length scale and the amplitude, \ie, $\lambda \, \Delta \rho =
0.01 \, \bar{\rho} \, L^{\mathrm{osc}}_{\mathrm{matter}}$, which is
indicated as a vertical line in the left plot. In both plots, the length
scale is fixed to be $5 \, \mathrm{km}$ (dotted curves), $500 \,
\mathrm{km}$ (dashed curves), and $2000 \, \mathrm{km}$ (solid curves),
respectively.
}
\end{figure*}
This can be seen in \fig~\ref{pert} (left plot), where the relative error 
$| \Delta P_{\mu e} |/P_{\mu e}^{\mathrm{ref}}\equiv|P_{\mu e}-P_{\mu
e}^{\mathrm{ref}}|/P_{\mu  e}^{\mathrm{ref}}$, coming from a
rectangular matter density perturbation, is plotted as a function of
the product of the length scale
$\lambda/L^{\mathrm{osc}}_{\mathrm{matter}}$ and the density contrast
$\Delta \rho/\bar{\rho}$.  For the oscillation parameters we choose
$\theta_{12} = 45^\circ$, $\theta_{23} = 45^\circ$, $\theta_{13} =
5^\circ$, $\Delta m_{32}^2 = 2.5 \cdot 10^{-3} \, \mathrm{eV}^2$,
$\Delta m_{21}^2 = 3.65 \cdot 10^{-5} \, \mathrm{eV}^2$, and
$\delta_{CP} = 0$, corresponding to the LMA solution with a value for
$\theta_{13}$ somewhat below the CHOOZ bound neglecting CP violating effects. 
The other parameters (and terms) are given and described in
the figure caption. The relative error is shown for several fixed
values of $\lambda \ll L^{\mathrm{osc}}_{\mathrm{matter}}$, \ie, only
the matter density is varied in such a way that the area of the perturbation
is unaffected for a fixed value on the horizontal axis. For the cut at
the vertical line this constant area of the matter density
perturbation is shown in the right plot, indicating that the
corresponding parameter value is already a quite pessimistic choice. 
Apparently, the curves in the left plot are approximately equal to each other
at least below the parameter value indicated by the vertical line and
especially for very small fixed values of $\lambda$. This means that the
relevant parameter for $\lambda \ll L^{\mathrm{osc}}_{\mathrm{matter}}$ is
the integral of the matter density perturbation, \ie, the product of
the length scale and the amplitude. Thus, we may expect that
interference effects, arising from non-commuting operators in the
Hamiltonian corresponding to different matter density layers, 
become irrelevant for very short length scales compared
with the oscillation length in matter. This can also be seen in the
analytical perturbation theoretical approach in \App~\ref{apppert}, where
the case of baselines much shorter than the oscillation length leads to first
order corrections $\propto \lambda \, \Delta \rho$ to the transition
probabilities, \ie, the product of the length scale and the amplitude is the
relevant parameter. Another result of the numerical analysis is that we do
not have to take into account isolated short scale perturbations, such as the
ones coming from the matter density contrast in, for example, a mine.
Estimating the length scale of such a perturbation to be shorter than $10 \,
\mathrm{km}$ and the relative density contrast to be of the order of $10 \%$,
we can read off a relative error much smaller than $1 \%$ from
\fig~\ref{pert}. Since the amplitude of the appearance probability $P_{\mu
e}$ is basically proportional to $\sin^2 2 \theta_{13}$ and the error in the
determination of $\theta_{13}$ is rather substantial \cite{Freund:2001ui}, we
will henceforth neglect very short isolated perturbations.

\section{A model for the matter density fluctuations}

\begin{figure}[ht!]
\begin{center}
\includegraphics[width=7cm]{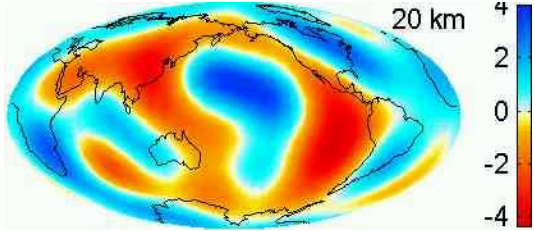}
\end{center}
\vspace*{-1cm}
\caption{\label{EarthMap} Percentage fluctuation in the Earth matter density
at a depth of $20 \, \mathrm{km}$ from seismic wave measurements
\cite{ekst98} (reprinted from \Ref~\cite{Pana}).}
\end{figure}

In this section, we will construct a model for the fluctuations
in the Earth matter density. Figure~\ref{EarthMap} shows the percentage
fluctuations in the Earth matter density at a depth of $20 \, \mathrm{km}$
below the Earth's surface obtained from seismic wave measurements. 
Though a neutrino beam traverses a large range of different depths, one may
take this figure as an estimate of the characteristic length scales and
amplitudes involved in the problem. It suggests length scales of
the order of some thousands of kilometers and amplitudes of the order of $\pm
4\%$, whereas at greater depths one can indeed have somewhat larger amplitudes.
In addition, the length scales and amplitudes do not seem to vary too much
around their average absolute values. This also implies that the
transition regions between negative and positive amplitudes are quite
short compared to the overall structure. So why not simply use these
measurements in neutrino physics, instead of discussing uncertainties
in the Earth matter density? First, these measurements contain some
averaging as well as uncertainties in the equation of state of the Earth matter
density profile from the seismic wave velocity profile (see, \eg,
\Refs~\cite{Jean86,Jean90}). Second, different groups obtain different results
\cite{Pana}, which are, however, not qualitatively so much different
with respect to the parameters we will identify below. 

\begin{figure*}[ht!] 
\begin{center}
\includegraphics[height=10cm]{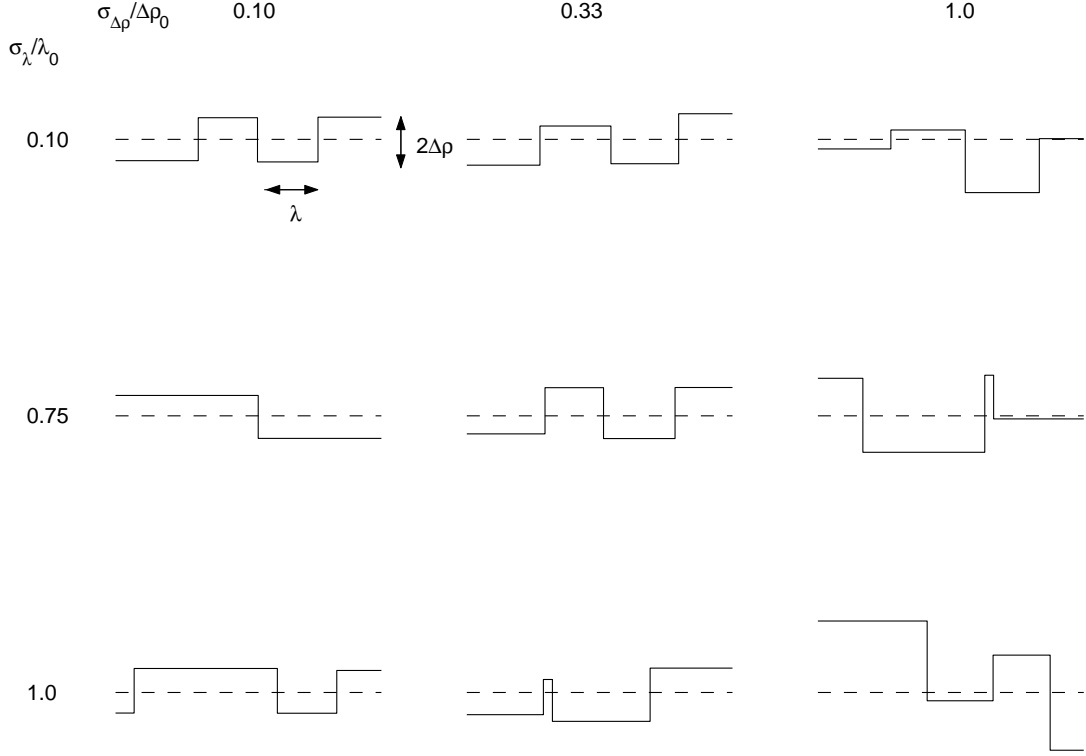} 
\end{center}
\vspace*{-1cm}
\caption{\label{profile_ex} Sample profiles for a fixed average length scale
$\lambda_0$ and a fixed average amplitude $\Delta \rho_0$, corresponding
to the values in \eq~(\ref{standardset}) and a baseline length of $L =
7400 \, \mathrm{km}$. The standard deviations are varied relative to
the absolute values of the length scale and the amplitude.}
\end{figure*}

In order to model these fluctuations realistically and investigate the 
dependence of the relevant parameters on the neutrino oscillation
transition probabilities, we use a step-function approach, varying the
absolute value of the amplitude $\Delta \rho >0$ and the length scale
$\lambda>0$ around some average values $\Delta \rho_0>0$ and
$\lambda_0>0$ at random. For the random variation 
we choose a Gaussian distribution with standard deviations
$\sigma_{\lambda}$ and $\sigma_{\Delta \rho}$, respectively, which is
truncated at zero. Figure~\ref{profile_ex} shows some sample profiles
for different values of these standard deviations and for fixed
average values.
Comparing this model with \fig~\ref{EarthMap}, we can then estimate a
set of realistic parameters as
\begin{subequations}
\begin{eqnarray} 
\lambda_0 & \sim & 2000 \, \mathrm{km},\\
\sigma_{\lambda} & \sim & 1500 \, \mathrm{km} = 0.75 \, \lambda_0,\\
\Delta \rho_0 & \sim & 3 \% \, \bar{\rho},\\
\sigma_{\Delta \rho} & \sim & 1 \% \, \bar{\rho} = 1/3 \, \Delta \rho_0,
\end{eqnarray}
\label{standardset} 
\end{subequations}
where $\bar{\rho} = 3 \, \mathrm{g/cm^3}$ is the average matter density.

\section{Numerical analysis}

For physical reasons we do not expect surprises in certain regions of the
parameter space. Since we cannot show the results for the whole
parameter space in $\lambda_0$, $\Delta \rho_0$, $\sigma_{\lambda}$, and
$\sigma_{\Delta \rho}$ simultaneously, we will now systematically investigate
the dependence on some parameters by keeping the other ones fixed.
One could do this by showing either the absolute errors in the
appearance probabilities coming from matter
density fluctuations $|\Delta P_{\alpha \beta} |$, or the relative errors
$|\Delta P_{\alpha \beta} |/P_{\alpha \beta}^{\mathrm{ref}}$. For the neutrino
channel $\nu_{\mu} \rightarrow \nu_{e}$, the transition probabilities are in
most regions relatively large. In this case, the relative errors are quite
meaningful and are usually some percent of the total probabilities. Depending
on the parameters they can sometimes even be larger than $10 \%$. However, for
the antineutrino channel $\bar{\nu}_{\mu} \rightarrow \bar{\nu}_{e}$ the
absolute probabilities in the denominators of the relative errors are rather
small. Therefore, it turns out that the relative errors are not very
sensible in this case. Comparing plots for the absolute and relative
errors and mainly focusing on the qualitative behavior, we thus decided to
show only the absolute error plots. Nevertheless, the fact that the absolute
values of these errors are rather small does not mean that they are small
compared to the transition probabilities. For the simulations, a large
number of matter density profiles was created at random and the relative
error was averaged over all computations with these profiles.

\begin{figure}[ht!]
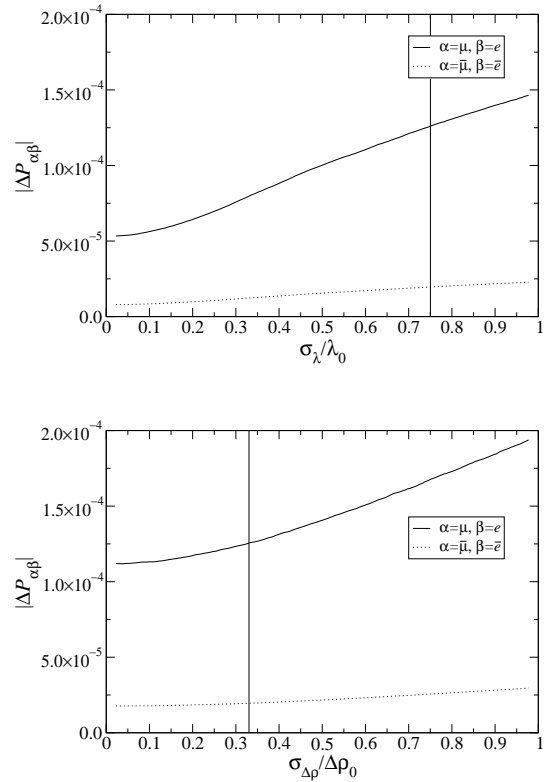

\begin{center} 
\includegraphics[bb=75 -35 645 690,angle=-90,width=7cm]{sl_plot_r}
\includegraphics[bb=75 -35 585 690,angle=-90,width=7cm]{sdr_plot_r}
\end{center}
\vspace*{-1cm} 
\caption{\label{sigma} The absolute errors in the appearance
probabilities $P_{\mu e}$ and $P_{\bar{\mu} \bar{e}}$, averaged over 
$20000$ random matter density profiles, with the values
$\lambda_0 = 2000 \, \mathrm{km}$ and $\Delta \rho_0 = 3 \% \cdot  3 \,
{\rm g/cm^3}$ from \eq~(\ref{standardset}). The errors are plotted as
functions of $\sigma_{\lambda}/\lambda_0$ for the fixed value $\sigma_{\Delta
\rho} = 1/3 \, \Delta \rho_0$ (upper plot) and $\sigma_{\Delta \rho}/\Delta
\rho_0$ for the fixed value $\sigma_{\lambda} = 0.75 \, \lambda_0$ (lower
plot), respectively [\cf, \eq~(\ref{standardset})]. The oscillation parameter
values are chosen as described in the caption of \fig~\ref{pert}.} 
\end{figure}

Figure~\ref{sigma} shows the absolute errors in the appearance probabilities
for the neutrino channel, $P_{\mu e}$, and the antineutrino channel, 
$P_{\bar{\mu} \bar{e}}$, plotted as functions
of $\sigma_{\lambda}$ for a fixed value of $\sigma_{\Delta \rho}$ and
$\sigma_{\Delta \rho}$ for a fixed value of $\sigma_{\lambda}$, respectively.
For the fixed parameter values we choose, if not otherwise noted, the
values from \eq~(\ref{standardset}). 
In all of our plots, the vertical
lines correspond to the parameter values in \eq~(\ref{standardset}), \ie,
the same point in the multi-dimensional parameter space. 
Since the curves in \fig~\ref{sigma} are slowly varying close to the
vertical lines, we will further on take the respective values from
\eq~(\ref{standardset}) for the standard deviations. In general, the absolute
errors are growing with larger fluctuations in the length scale and
the amplitude. This means, in our model,  
that for more irregular matter density fluctuations
we obtain larger errors in the transition probabilities.
A possible explanation could be the zero-truncations of the Gaussian
distributions, which means that for large standard deviations
the average values are shifted.
For antineutrinos the absolute errors are in all cases much 
smaller than for neutrinos, because for antineutrinos matter effects are, in
general, much smaller for energies larger than a few GeV (no resonance
effects). 
 
\begin{figure}[ht!]
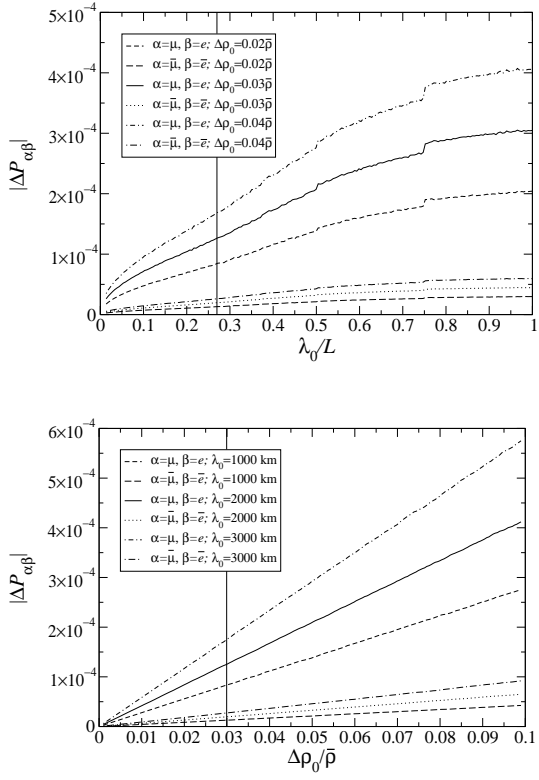
 
\begin{center} 
\includegraphics[bb=75 -35 645 690,angle=-90,width=7cm]{l0_plot_r}
\includegraphics[bb=75 -35 585 700,angle=-90,width=7cm]{dr_plot_r}
\end{center}  
\caption{\label{scales} The absolute errors in the appearance
probabilities $P_{\mu e}$ and $P_{\bar{\mu} \bar{e}}$, averaged over
$20000$ random matter density profiles, with the standard
deviation values $\sigma_{\lambda} = 0.75 \, \lambda_0$ and
$\sigma_{\Delta \rho} = 1/3 \, \Delta \rho_0$, corresponding to
\eq~(\ref{standardset}). The errors are plotted as functions of the baseline
fraction $\lambda_0/L$ for three different values of $\Delta \rho_0$ (upper
plot) and the average matter density fraction $\Delta \rho_0/\bar{\rho}$ for
three different values of $\lambda_0$ (lower plot), respectively. Note
that the standard deviations scale linearly with the absolute
average values. The other parameter values are chosen as described in
the caption of \fig~\ref{pert}.}
\end{figure}

Next, let us investigate the dependence of the absolute errors on the length
scale $\lambda_0$ and the amplitude $\Delta \rho_0$. Here we choose
the relative standard deviations to be $\sigma_{\lambda} = 0.75 \,
\lambda_0$ and $\sigma_{\Delta \rho} = 1/3 \,
\Delta \rho_0$, corresponding to \eq~(\ref{standardset}) for the appropriate
$\lambda_0$ and $\Delta \rho_0$.
The result of this analysis is shown in \fig~\ref{scales}, from which it can be
seen that the absolute error is essentially proportional to the length scale
as well as the amplitude of the fluctuations.
Indeed, this result is, with respect to the average length scale and
amplitude, very similar to what we obtained from a single perturbation
in the Earth matter density. In the latter case, we observed that the
product of the length scale and amplitude determines the error in the
probabilities. Thus, fixing one of these two parameters gives a linear
dependence on the other one. This is, of course, only true for a
single perturbation without interference effects, which means that
corrections to linearity have to be taken into account in our more
general model. One such interesting correction is the bumps in the
upper plot. Since in our model the number of steps $n$, in which the
matter density profile has been divided into, depends on the length
scale, for a small number of steps $n$ the (average) transition $n \rightarrow
n-1$ can be seen as a bump in the absolute errors. For large $n$ the relative
contribution of this effect becomes negligible.

\begin{figure}[ht!]
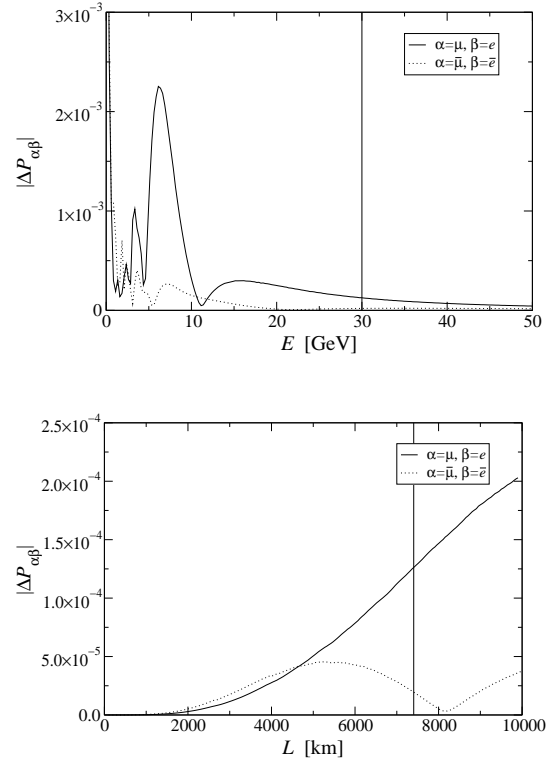

\begin{center} 
\includegraphics[bb=75 -35 645 700,angle=-90,width=7cm]{E_plot_r}
\includegraphics[bb=75 -35 585 715,angle=-90,width=7cm]{L_plot_r}
\end{center}
\vspace*{-1cm}
\caption{\label{enbaseline} The absolute errors in the appearance
probabilities $P_{\mu e}$ and $P_{\bar{\mu} \bar{e}}$, averaged over 
$20000$ random matter density profiles, with the parameter values from
\eq~(\ref{standardset}). The errors are plotted as functions of the energy
$E$ for a baseline of $L=7400 \, \mathrm{km}$ (upper plot) and the
baseline $L$ for an energy of $E=30 \, \mathrm{GeV}$ (lower plot). The
other parameter values are chosen as described in the caption of
\fig~\ref{pert}.}
\end{figure}

Finally, taking the parameter values as given in \eq~(\ref{standardset}), the
absolute errors in the appearance probabilities $P_{\mu e}$ and $P_{\bar{\mu} 
\bar{e}}$ are plotted as functions of energy and baseline in
\fig~\ref{enbaseline}. 
{}From the upper plot we can observe that the absolute errors are
rather small for large energies ($E \gtrsim 10 \, \mathrm{GeV}$)
and so are the relative errors at least for neutrinos. Since for a
neutrino factory the energy spectrum for muon neutrinos is
peaked at the maximum energy \cite{Geer:1998iz,Barger:1999fs}, the low-energy
contributions contain less statistical information. This point supports our
focus on high energies in most of the plots. 

As far as the baseline dependence in the lower plot of
\fig~\ref{enbaseline} is concerned, the absolute errors are vanishing for small
baselines, because matter effects become, in general,
negligible. Although the neutrino and antineutrino curves look quite
different in this plot, the basic principle of the shapes is the
same. They are both periodic functions of the baseline, as can easily
be seen for the antineutrinos, and the difference is only due to
different period and amplitude.

\section{Summary and conclusions}

In this paper, we first investigated the effects of a single
perturbation in the average matter density of the Earth (especially the
Earth's mantle). We then introduced a model for the description of
random matter density fluctuations in the Earth's mantle, which is
based on the observations obtained from geophysical
measurements. Finally, we used this model to analyze the dependence of 
the absolute errors in the appearance probabilities $P_{\mu e}$ and
$P_{\bar{\mu} \bar{e}}$, especially important for long baseline neutrino
factory CP measurements, on our model parameters.

We observed that, in particular for neutrinos, the absolute error in the
appearance probability for random matter density fluctuations can be
quite substantial, corresponding to relative errors of some few
percent. The error is essentially directly proportional to the product
of the amplitude and length scale of a single perturbation or the
random fluctuations. 
Furthermore, for short baselines $L \lesssim 1000 \, \mathrm{km}$,
the errors are vanishing for both neutrinos and antineutrinos 
which means that matter density fluctuations can in this case be
regarded as second order effects compared to the small matter effects.

Finally, we comment on the three possible scenarios suggested in the
summary of \Ref~\cite{Geller:2001ix}:
\begin{enumerate}
\item
The uncertainty of present density models poses no significant problems.
\item
Moderate reduction of the uncertainty, through more detailed analysis of data,
is required.
\item
Significant reduction of this uncertainty, by conducting a large scale
campaign of geophysical observations, is required.
\end{enumerate}
{}From our analysis, scenario (2) best fits our conclusions. In our
calculations with randomly generated matter density profiles, we assumed that
the fluctuations are completely unknown and we obtained relative
errors of the order of magnitude of some few percent. Comparing the
results of the measurements of the PREM profile corrections of
different groups \cite{Pana} indicates that there is not yet
sufficient agreement on the data. Moderate reduction of the uncertainty by
a more detailed analysis of data should help to settle this problem
for neutrino physics.

\begin{appendix}

\section{A perturbation theoretical approach to a single perturbation in the
Earth matter density} 
\label{apppert}

In perturbation theory, we can show that for a single perturbation with
constant amplitude $\Delta A \equiv \pm 2 \, \sqrt{2} \, G_F \, E \, \Delta n_e
\propto \Delta \rho$ in the average Earth matter density $A \propto
\bar{\rho}$, the first order perturbation term is proportional to the ``area''
$\Delta S \equiv \lambda \, \Delta A \propto \lambda \, \Delta \rho$ of the
perturbation, \ie, the length scale $\lambda$ times the amplitude $\Delta A$.

The Hamiltonian for the propagation of the neutrinos is
\begin{equation}
\mathcal{H}(r)=\mathcal{H}_{0}+A \, \mathcal{K} + \Delta A \,
\mathcal{K} \equiv \mathcal{H} + \Delta A \, \mathcal{K},
\label{hamilton}
\end{equation}
where $\mathcal{K} \equiv | \nu_e \rangle \langle \nu_e |$ is the projector
onto the flavor state $| \nu_e \rangle$. Let $\mathcal{V}(r)=e^{-i
\mathcal{H}r} =\sum_{a=1}^{3} e^{-i\xi_{a}r} \, \mathcal{P}_{a}$ be the
unperturbed evolution operator, where $\mathcal{P}_{a} \equiv | \nu_a \rangle
\langle \nu_a |$ is the projector onto the mass eigenstate $| \nu_a \rangle$
with the eigenvalue $\xi_a$ in matter. Then the evolution operator of the full
evolution equation can, to first order in perturbation theory, be written as 
\begin{equation}         
\mathcal{U}(r) \simeq  \mathcal{V}(r) -i \mathcal{V}(r) \, \Delta A
\sum_{a=1}^3 \sum_{b=1}^3 f_{ab} \, \mathcal{P}_{a} \, \mathcal{K} \,
\mathcal{P}_{b},   
\label{perturbation} 
\end{equation}
where
\begin{equation}
f_{ab} \equiv  \lambda \, e^{i(\xi_{a}-\xi_{b}) r_{0}} \, \frac{\sin
\left(\xi_{a}-\xi_{b} \right) \frac{\lambda}{2}}{\left(
\xi_{a}-\xi_{b} \right) \frac{\lambda}{2}}
\label{fab}
\end{equation}
and $r_{0}$ is the position of the center of the fluctuation.
For a fluctuation length $\lambda$, which is much shorter than the
oscillation length, \ie, $(\xi_{a}-\xi_{b}) (\lambda/2) \ll 1$, the last
factor in \eq~(\ref{fab}) is approximately equal to unity. We then obtain
\begin{equation}
f_{ab} \simeq \lambda \,  e^{i(\xi_{a}-\xi_{b})r_{0}}.
\label{fab2}
\end{equation}
With this result it is easy to see that the transition probability $P_{\alpha
\beta}(r)$ at the position $r$ can be written as
\begin{equation}
P_{\alpha \beta}(r) \simeq P^{0}_{\alpha \beta}(r) + 2 \, \Delta S \,
\Im [X^{*}(r)Y(r)],
\label{prob}
\end{equation}
where $P^{0}_{\alpha \beta}(r)$ is the unperturbed transition 
probability in constant matter density,
$X(r) \equiv \langle \nu_{\beta} | \mathcal{V}(r)| \nu_{\alpha} \rangle$, and
$Y(r) \equiv \langle \nu_{\beta} | \left( \mathcal{V}(r_{0}-r)
\right)^{\dagger} |\nu_e 
\rangle \langle  \nu_e |\mathcal{V}(r_{0})| \nu_{\alpha} \rangle$.
This shows that the perturbary contribution to the transition 
probability is proportional to $\Delta S$, and therefore, largely
independent of the form of the perturbation. Note that perturbation theory
only holds for $\Delta S$ small compared to $\langle \mathcal{ H}
\rangle \, r \simeq A \, r$, \ie, $\Delta A/A \ll r/\lambda$.

\end{appendix}

\section*{Acknowledgments}

T.O. and W.W. greatly acknowledge the support for the visit at KTH -
SCFAB, where a large part of this work was carried out, as well as for
the warm hospitality.

This work was supported by the Swedish Foundation for International
Cooperation in Research and Higher Education (STINT) [T.O.], the 
Wenner-Gren Foundations [T.O.], the ``Sonderforschungsbereich 375 f{\"u}r
Astro-Teilchenphysik der Deutschen Forschungsgemeinschaft'' [T.O. and
W.W.], and the Swedish Natural Science Research Council (NFR), Contract 
No. F 650-19981428/2001 [H.S.].

\end{document}